# DFT Study of Nitroxide Radicals.

## 1. Effects of solvent on structural and electronic characteristics of 4-amino-2,2,5,5-tetramethyl-3-imidazoline-N-oxyl

Larissa N. Ikryannikova[1*], Leila Yu. Ustynyuk[1], Alexander N.Tikhonov[2**]

[1]*Department of Chemistry and* [2]*Department of Physics, Moscow State University, Moscow, Russia*

[*]Corresponding author. Tel: (+7)-(095)-9393685, fax: (+7)-(095)-9394575,

e-mail address: larisa@kge.msu.ru (L.N. Ikryannikova).

[**]Corresponding author. Tel: (+7)-(095)-9392973, fax: (+7)-(095)-9391195,

e-mail address: an_tikhonov@newmail.ru (A.N.Tikhonov).

## Abstract

Imidazoline-based nitroxide radicals are often used as spin probes for medium acidity and polarity in different systems. In this work, using the density functional theory (DFT) approach, we have studied how physico-chemical characteristics (geometry, atomic charges and electron spin density distribution) of pH-sensitive spin label 4-amino-2,2,5,5-tetramethyl-3-imidazoline-N-oxyl (ATI) depend on protonation and aqueous surroundings. Our calculations demonstrate that ATI protonation should occur at the nitrogen atom of the imidazoline ring rather than at the amino group. Protonation of ATI leads to a decrease in a spin density on the nitrogen atom of the nitroxide fragment >N-O$^{\bullet}$. For simulation of ATI hydration effects, we have constructed a water shell around a spin label molecule by means of gradual (step-by-step) surrounding of ATI with water molecules ($n_{\mathrm{H_2O}}$ = 2-41). Calculated spin density on the nitrogen atom of the nitroxide fragment increased with an extension of a water shell around ATI. Both protonation and hydration of ATI caused certain changes in calculated geometric parameters (bond lengths and valence angles). Investigating how structural and energy parameters of a system ATI–(H$_2$O)$_n$ depend on a number of surrounding water molecules, we came to the conclusion that a hydrogen-bonded cluster of $n_{\mathrm{H_2O}} \geq 41$ water molecules could be considered as an appropriate model for simulation of ATI hydration effects.

## 1. Introduction

Nitroxide radicals are widely used as spin probes in different fields of chemistry, physics and biology to investigate structural and dynamic properties of biopolymers, membranes and different nanostructures[1-12]. Nitroxide radicals also serve as molecular probes for oxygen and its active forms[13-16]. Practical usage of spin probes is based on sensitivity of their electron paramagnetic resonance (EPR) spectra to molecular mobility and environment (medium polarity, pH, microviscosity and ordering of solvent molecules)[1-3,17-20]. A development of adequate theoretical approaches to determination of structural and EPR characteristics of spin probes still remains as one of the most actual tasks of computational chemistry[21-22]. The most informative EPR parameters of nitroxide radicals are the g-tensor and hyperfine splitting tensor, which are determined by the interaction of the unpaired electron with the nitrogen nucleus of the $>N-O^{\bullet}$ fragment. Quantum chemical calculations of spectral parameters of nitroxide radicals is complicated, in particular, by the necessity of accounting for interactions of nitroxide radicals with solvent molecules. It is well known that polarity of a solvent strongly effects on the isotropic splitting constant, $a_{iso}(N)$, of nitroxide radicals[1-9].

Several attempts to take into account the solvent effects were undertaken in a number of theoretical works[23-31]. One of traditional approaches to this problem is based on the use of a polarizable continual model (PCM)[24-26]. This model considers a solute molecule placed inside the cavity surrounded by a continual medium characterised by a dielectric constant $\varepsilon$. The cavity may be specified, e.g., as a set of overlapping spheres placed at atoms of a solute molecule. This approach, however, neglects specific interactions between the solvent and spin probe molecules. A further development of this model was suggested by Barone and collaborators who combined the PCM approach with a direct consideration of spin probe interactions with the 'local aqueous surroundings' (two or three water molecules placed in the vicinity of the $>N-O^{\bullet}$ fragment)[27-31]. However, even in this case additional amendments were needed to get satisfactory values of $a_{iso}$. Therefore, it seems tempting to consider more realistic model for hydrated nitroxide radicals that would take into account the interaction of a spin label molecule with a large number of solvent molecules. This is the motivation for the present work. One of important issues is to construct a hydrogen-bonded network of water molecules that might be considered as an appropriate model for a complete first hydration sphere surrounding a spin label molecule.

Among a great variety of different spin labels, imidazoline-based nitroxide radicals are of special interest for biological and chemical applications[4,9]. Spin probes of this kind are used as pH-sensitive molecular probes, because the EPR spectra of protonated and deprotonated spin labels are characterized by different hyperfine splitting constants and, therefore, can be easily detected by the EPR spectroscopy method[32-34]. In this work, we have studied the influence of an aqueous surroundings on physico-chemical characteristics (geometry, energy, distribution of charges and electron spin density on atoms) of pH-sensitive spin probe 4-amino-2,2,5,5-tetramethyl-3-imidazoline-N-oxyl (ATI) (Fig.1). Using the density functional method, we have calculated these characteristics for ATI molecules surrounded by various numbers of water molecules. We wanted to know how a stepwise extension of a water cluster would influence on structural and electron properties of ATI molecule. Thus we could evaluate a minimal size of a water cluster that might be considered as a model for a first hydration sphere around ATI.

## 2. Computational strategy, methods and details

Our computational strategy stems from the recent developments of advanced DFT methods for low cost computations of rather large (more than 100-150 atoms) molecular systems[21]. All calculations were carried out with the use of the original program PRIRODA developed by D.N. Laikov[35]. This program allows relatively fast DFT calculations to be performed for large molecular systems without the loss of computation accuracy. Therefore, we could perform our calculations on powerful PC during a moderate period of time.

All our calculations were performed with exchange-correlation functionals PBE and BLYP. These functionals include electron density gradients. Functional PBE is one of the best generalized gradient approximation (GGA) for the exchange-correlation functional by Perdew, Berke, and Ernzerhof[36]. Functional BLYP is widely used for calculations of spin density in nitroxide radicals[21,37]. Gaussian-type basis sets and the expansion of electron density in the auxiliary basis were used for the solution of Kohn-Sham equations[39]. We used two types of orbital and auxiliary basis sets (B1 and B2). Contraction patterns for these basis sets are presented in Table 1. Geometries of all the systems considered in this work were optimized with the use of basis B1. Full optimization of geometry was performed using analytical gradients. For gas phase calculations, it was followed by analytical calculations of the second derivatives of energy with respect to coordinates in order to characterize the nature of the resulting stationary points (minima or saddle points) on the potential energy surface. Basis B2 was used for single point calculations of energy and spin densities performed after preliminary geometry optimization with the use of basis B1.

In order to construct an aqueous sphere around a spin label protonated at the position of the ring nitrogen N4, ATI(H$^+$), we started with a small system (ATI(H$^+$) + 2H$_2$O) and then expanded step-by-step a size of a water cluster until a complete hydrogen-bonded water network surrounding ATI(H$^+$) was formed (Fig.2). On the first stage, one water molecules was placed near the nitroxide fragment >N–O$^{\bullet}$, while another one was positioned near the protonated nitrogen atom of the imidazoline ring. After the geometry of this system was optimised, other two H$_2$O molecules were added and then the geometry of a new system was optimised again. Such a procedure of the stepwise expansion of a water cluster was repeated several times, until a complete hydrogen-bonded aqueous shell around ATI(H$^+$) was constructed. Geometry of each cluster ATI(H$^+$)−(H$_2$O)$_n$ was optimised using the B1 basis set. Optimization of each molecular system was performed without any symmetry constraints. The convergence limit used for geometry optimization was 10$^{-7}$.

On the basis of geometry of optimised ATI(H$^+$)−(H$_2$O)$_{41}$ cluster we constructed three other model systems: (i) hydrated ATI molecule with protonated amino group, ATI(-NH$_3^+$)−(H$_2$O)$_{41}$; (ii) deprotonated ATI molecule surrounded with 41 H$_2$O molecules, ATI−(H$_2$O)$_{41}$ and (iii) deprotonated ATI molecule surrounded with one H$_3$O$^+$ and 40 H$_2$O molecules, ATI-(H$_3$O$^+$)(H$_2$O)$_{40}$. In the first case, the proton bound to the nitrogen atom N4 of the imidazoline ring was transferred to the amino group and then geometry of a cluster ATI(-NH$_3^+$)−(H$_2$O)$_{41}$ was re-optimised. In the second case, the proton bound to N4 atom was removed and then geometry of deprotonated system ATI−(H$_2$O)$_{41}$ was re-optimised. In the latter case, the proton was transferred to the nearest water molecule placed in the vicinity of N4 atom and then geometry of protonated system ATI-(H$_3$O$^+$)(H$_2$O)$_{40}$ was re-optimised.

For comparison, the solvation effects were also estimated using the IPCM model. This option is incorporated into the program PRIRODA[35].

Partitioning of charges and electron spin densities on the atoms of each optimized system was calculated according to the reference[38]. The isotropic hyperfine coupling constant $a_{\text{iso}}(\text{N2})$, which measures the contacts Fermi interaction between the unpaired electron and the nitrogen nucleus of the >N−O$^\bullet$ fragment, was calculated as

$$a_{\text{iso}}(\text{N2}) = \frac{8}{3}\pi \cdot g_{\text{e}}\beta_{\text{e}}g_{\text{n}}\beta_{\text{n}}\rho(\vec{r}_{\text{N}})\,. \tag{1}$$

Here, $\rho(\vec{r}_{\text{N}})$ represents the electron spin density at the nitrogen nucleus, $g_{\text{e}}$ is the free-electron g-factor, $\beta_{\text{e}}$ is the Bohr magneton, $g_{\text{n}}$ and $\beta_{\text{n}}$ are the nuclear g-factor and nucleus magneton, respectively. For calculations of $a_{\text{iso}}(\text{N2})$ values, we used B2 or B1 basis sets, relying on a system geometry optimized previously with B1 basis set.

To the best of our knowledge, there are no literature data regarding structural parameters of ATI molecule. Therefore, to verify the applicability of the computation methods used in this work, we compared results of our calculations performed for one of the most simple nitroxides, 4 hydroxy-2,2,6,6-tetramethylpiperidine-1-oxyl (TEMPOL), with the literature data - experimental and computed structural parameters of TEMPOL[39]. Table 2 lists geometry parameters of two possible conformers of TEMPOL computed in our work at PBE/B1 and BLYP/B1 levels. One can see that both approaches give results which are in an agreement with calculated and experimental structural parameters of TEMPOL published earlier[39].

## 3. Results and discussion

### 3.1. Sites of ATI protonation.

There are two potential protonation sites in ATI molecule – the nitrogen atom N4 of the imidazoline ring and the amino group (Fig. 1). To compare the probabilities of ATI protonation at these sites, we calculated the energies of ATI molecules protonated either at the N4 position (denoted as ATI(H$^+$)) or at the amino group (denoted as ATI(-NH$_3^+$)). According to our calculations, the ring nitrogen N4 is characterised by higher affinity to proton than to the amino group. For ATI molecule in the gas phase (Fig. 3A), the energy difference is $\Delta E_{gas} = E_{ATI(-NH_3^+)} - E_{ATI(H^+)} = 31.4$ kcal/mol. The energy of the transition state $[ATI-H^+]^{\#}$ of the interconversion process ATI(H$^+$) $\leftrightarrow$ ATI(-NH$_3^+$) is higher than the energies of ATI(H$^+$) and ATI(−NH$_3^+$) by 55.2 and 23.8 kcal/mol, respectively (Fig. 3A). Therefore, a direct proton transfer between the amino group and N4 atom should be hindered kinetically.

Computations for hydrated ATI (clusters ATI(H$^+$)–(H$_2$O)$_{41}$ and ATI(-NH$_3^+$)–(H$_2$O)$_{41}$,) also demonstrate that the ring nitrogen N4 has higher affinity to proton than the amino group (Fig. 3B). In this case, the energy difference is $\Delta E_{water} = E_{ATI(-NH_3^+)-(H_2O)_{41}} - E_{ATI(H^+)-(H_2O)_{41}} = 19.1$ kcal/mol, which is smaller than $\Delta E_{gas}$. When ATI molecule is dissolved in water, the proton transfer between two protonation sites (N4 and N7 atoms) could occur via a hydrogen-bonded network of water molecules surrounding ATI. In order to evaluate the possibility of this mechanism, we calculated an energy of a system ATI–H$_3$O$^+$(H$_2$O)$_{40}$ consisted of deprotonated ATI molecule surrounded by a water cluster H$_3$O$^+$(H$_2$O)$_{40}$. This system should simulate one of intermediate states on the way of the proton migration between the amino group and N4 atom along a hydrogen-bonded water network. The energy diagram shown in Fig. 3B demonstrates that the efficient activation energies for the proton transfer processes ATI(-NH$_3^+$)–(H$_2$O)$_{41}$ $\leftrightarrow$ ATI–H$_3$O$^+$(H$_2$O)$_{40}$ and ATI(H$^+$)–(H$_2$O)$_{41}$ $\leftrightarrow$ ATI–H$_3$O$^+$(H$_2$O)$_{40}$ are $E'_{a1} = 3.3$ and $E'_{a2} = 22.4$ kcal/mol, respectively. It is reasonable to assume that the relatively low activation energy (3.3 kcal/mol) cannot be considered as an obstacle to the downhill proton transfer from the amino group to the ring nitrogen N4. Therefore, we conclude that the proton binding would occur predominantly at the N4 atom of the imidazoline ring. Our conclusion is in agreement with experimental data on the EPR study of ATI spectra in water solution[34]. Bearing in mind this result, we consider below only the ATI(H$^+$) form of protonated ATI molecule.

### 3.2. Geometry parameters of ATI molecule.

In this section, we analyse how protonation and hydration of ATI molecule effect on the ATI geometry. Geometry parameters of ATI molecule (bond lengths, angles and torsions) presented in Table 3 were calculated for protonated and deprotonated ATI molecules in the gas phase or ATI molecule surrounded by a water shell of 41 $H_2O$ molecules. The following structural parameters were analysed: 1) a bond length of the nitroxide fragment $l_{N-O}$; 2) an angle $\alpha$ which characterises a deviation of the nitroxide fragment >N-O$^\bullet$ from the plane determined by atoms N2, C3 and C6, 3) a dihedral angle $\beta$ which is the complement to the angle formed by planes N2–C3–C6 and C3–N4–C5–C6 (Fig. 1). Correctness of the choice of parameter $\beta$ is proved by the fact that four atoms of the imidazoline ring, C3, N4, C5 and C6, are situated practically in one plane. Really, almost in all the cases presented in Table 3, the torsion angle C3-N4-C5-C6 is less than one degree.

### 3.2.1. Effects of ATI protonation.

Table 3 lists results of calculations that clearly demonstrate that protonation of N4 atom of the imidazoline ring causes certain changes in the ATI geometry. In both cases (ATI in the gas phase and hydrated ATI), the protonation leads to a certain shortage of the N2-O1 bond length and causes an elongation of the double bond between N4 and C5 atoms. Bond angles also change with ATI protonation. The most interesting issue is a reversal of the signs of all calculated dihedral angles (except the C3-N4-C5-C6 one) upon the ATI protonation.

### 3.2.2. Effects of ATI hydration.

According to our calculations (Table 3), water surroundings influence on geometry parameters of both protonated and deprotonated forms of ATI. In both cases, the ATI hydration causes an elongation of the bond lengths O1-N2 and N4-C5.

In order to evaluate a minimal size of a water cluster that might be considered as an appropriate model for a complete first hydration shell around ATI, consider how structural parameters of ATI($H^+$) change with its stepwise surrounding by water. Fig. 4A shows the plot of a bond length $l_{N-O}$ versus a number of water molecules ($n_{H_2O}$) in a cluster ATI($H^+$)–($H_2O$)$_n$. Parameter $l_{N-O}$ increases with the rise of $n_{H_2O}$, reaching approximately the steady level $l_{N-O} \approx 1{,}28$ Å (at $n_{H_2O} = 41$). It is noteworthy that the main influence of water occurs after the addition of relatively small amounts of $H_2O$ molecules ($n_{H_2O} \leq 8$). With a further increase in a number of water molecules, we obtained insignificant variations of a bond length $l_{N-O}$ (in the interval $10 \leq n_{H_2O} \leq 41$). These variations reflect the discreteness of the system, when a number of water molecules is not high enough to form a complete hydration shell around the ATI molecule.

The conclusion about the water-induced structural distortions of the ATI geometry can be also illustrated by the plots of the valence angles $\alpha$ and $\beta$ versus a number of water molecules in a cluster ATI(H$^+$)–(H$_2$O)$_n$ (Figs. 4B and 4C). For ATI(H$^+$) molecule in the gas phase, we obtained that $\alpha < 0$ and $\beta > 0$. Addition of water changed the signs of these angles ($\alpha > 0$ and $\beta < 0$). Similarly to structural parameter $l_{N-O}$ (Fig. 4A), these changes can be induced by relatively small amounts of water molecules. We also obtained certain variations of parameters $\alpha$ and $\beta$ in the interval $10 < n_{H_2O} < 41$.

Water-induced structural changes in ATI molecule, which are characterised by changes in the signs of torsion angles, indicate that this molecule could have two possible conformations (Fig. 5). For protonated ATI(H$^+$) molecule in the gas phase, the energy difference calculated for these conformers is very small ($\leq 11$ cal/mol). The energy barrier separating the two conformational states is ac. 100 cal/mol. Fig. 4B demonstrates that the sign of an angle $\alpha$ oscillates with the extension of a water cluster. This suggests that in water solution ATI(H$^+$) could exist in two different conformations. In case of incomplete hydration shell, however, a rigorous interpretation of ATI structural changes is complicated by the lack of appropriate experimental data. Note that an aqueous cluster around ATI molecule could be considered as an adequate model for the hydration shell only when a further extension of this cluster would influence only insignificantly on the structural parameters of a spin label. According to our calculations (Fig. 4), this should be the case when $n_{H_2O} \approx 41$.

### 3.3. Energy characteristics of ATI-H$^+$–(H$_2$O)$_n$ clusters

In order to look for another criteria for the completeness of the first hydration sphere, we also analysed how an energy of a cluster ATI(H$^+$)–(H$_2$O)$_n$ depended on a number of water molecules in the system. To characterise the interaction of one water molecule with other molecules of a cluster ATI(H$^+$)–(H$_2$O)$_n$, we consider the energy parameter $\Delta E(n)$

$$\Delta E(n) = \frac{1}{n}[E(n) - E(0)] - E_0. \qquad (2)$$

Here, $E(0)$ is the energy of ATI in the gas phase, $E_0$ relates to the energy of isolated water molecule in the gas phase, and $E(n)$ is the energy of ATI(H$^+$)–(H$_2$O)$_n$. Thus, parameter $\Delta E(n)$ represents a mean energy of hydrogen bonds formed by one water molecule with other molecules in a system ATI(H$^+$)–(H$_2$O)$_n$.

Fig. 6 shows how parameter $\Delta E(n)$ depends on a number of water molecules surrounding ATI(H$^+$). The interaction of two H$_2$O molecules with ATI(H$^+$), when one water molecule forms the hydrogen bond with the O1 atom of the >N-O$^\bullet$ fragment and another one is placed near the N4 atom of the imidazoline ring, is characterised by the energy decrease $\Delta E(n) = -12.7$ kcal/mol. For a rather small number of water molecules ($2 < n_{H_2O} < 16$), parameter $\Delta E(n)$ oscillates around this level. When a number of water molecules becomes high enough ($n_{H_2O} \geq 26$), parameter $\Delta E(n)$ decreases up to the steady level $\Delta E(n) = -14.2$ kcal/mol [1]. This parameter remains practically unchanged with a further extension of a hydrogen-bonded water network around ATI ($26 \leq n_{H_2O} \leq 41$). Thus, along with the data presented in Section 3.2, we conclude that the water cluster of $n_{H_2O} \approx 40$ molecules could be considered as an appropriate model for simulation of a complete water sphere around ATI.

The solvation energy for protonated ATI molecule can be estimated as

$$E_{solv} = (E_{ATI(H^+)}^{gas} + E_{(H_2O)_n}) - E_{ATI(H^+)-(H_2O)_n}, \tag{3}$$

where $E_{ATI(H^+)}^{gas}$ is the energy of ATI(H$^+$) in the gas phase, $E_{ATI(H^+)-(H_2O)_n}$ is the energy of a cluster ATI(H$^+$)–(H$_2$O)$_n$, $E_{(H_2O)_n}$ is the energy of a pure water cluster (without a spin label) calculated after withdrawal of ATI(H$^+$) from a cluster ATI(H$^+$)–(H$_2$O)$_n$. Energy $E_{(H_2O)_n}$ was calculated without optimization of a water cluster geometry. For $n_{H_2O} = 41$, we obtained $E_{solv} = 94.4$ kcal/mol, which is about of two times higher than the solvation energy $E_{solv}^{IPCM} = 44.2$ kcal/mol calculated within the frames of IPCM approach (for the model parameter $\varepsilon_{H_2O} = 78.39$).

### 3.4. The influence of water environment on electronic characteristics of ATI.

In this section, we analyse how electronic characteristics of ATI (distributions of charges and spin density) depend on an aqueous surroundings of protonated and deprotonated forms of ATI. Figures 7 and 8 clearly demonstrate that water surroundings ($n_{H_2O} = 41$) causes redistribution of spin density and charges on the atoms of both protonated and deprotonated ATI. Below we consider results of our calculations, performed at different levels of theory, in more details.

### 3.4.1. Partial charges on atoms.

*Effects of protonation.* Fig. 7 shows that in both cases, dehydrated and hydrated ATI, the proton binding to N4 atom causes definite changes in partial charges on the atoms of a spin labels molecule. Protonation of ATI is accompanied with an increase of charges on positively charged atoms (O1, C3, C5, C6) and a concomitant decrease in charges of negatively charged atoms (O1, N4, N7). Charges on the atoms of methyl groups are effected by ATI protonation but only slightly.

---

[1] This value differs only slightly from the mean energy characterising the interaction of one water molecule with other in a pure water cluster formed of 27 water molecules. The formation of the network of hydrogen bonds between water

As we will see below (section 3.4.2), protonation-induced changes in charges on atoms are correlated with a certain redistribution of spin density in the nitroxide fragment of ATI molecule.

   *Effects of hydration.* Figure 7 shows that aqueous environment influences on charge distribution in both deprotonated and protonated forms of ATI (Fig. 7). The most significant changes occur in the nitroxide fragment N2-O1, as well as in the amino group and N4-H bond. Hydration of ATI and ATI($H^+$) molecule causes a rise in the positive charges on their N2 atoms, whereas the negative charges on O1 atoms become smaller. As concerns the N4-H bond and amino group, effects of water-induced redistribution of charges are different in cases of protonated and deprotonated ATI. For deprotonated form, negative charges of the nitrogen atoms N4 and N7 decrease after a spin label hydration. In contrast, the hydration of protonated molecule ATI($H^+$) results to a certain increase in the values of negative charges of N4 and N7 atoms. Positive charges of hydrogens in the amino group, as well as the positive charge of the hydrogen atom bound to the N4 atoms of the imidazoline ring, decrease with the ATI hydration.

   It is important to note that water-induced changes in charges on atoms cannot be attributed entirely to charge redistribution within a spin label molecule. Our calculations clearly demonstrate that water molecules surrounding ATI($H^+$) form hydrogen bonds with certain atoms of a spin label molecule (the amino group protons, N4-H group proton, and the oxygen atom of the nitroxide fragment). These hydrogen bonds provide a channel for a "spillover" of partial charges from protonated spin label to neighbouring water molecules. Actually, the total charge of ATI($H^+$) molecule, which equals to $Q_{ATI(H+)} = +1$ (a.u.) in the gas phase, decreases up to $Q_{ATI(H+)} = +0.45$ when ATI($H^+$) molecule has an aqueous environment ($n_{H_2O} = 41$). Remaining charge $Q_{(H_2O)_{41}} = +0.55$ is distributed over the water molecules of a cluster ATI($H^+$)-($H_2O$)$_{41}$. This result demonstrates that charged ATI($H^+$) molecule causes the polarisation of a surrounding water cluster. In the case of deprotonated (neutral) ATI molecule, as we could expect, there is no significant charge spillover to an aqueous cluster ($Q_{(H_2O)_{41}} \leq 0.02$).

   To analyse the charge redistribution effects in more details, let us consider the plots of charges on atoms O1 ($Q_{O1}$), N2 ($Q_{N2}$) and N4 ($Q_{N4}$) versus a number ($n_{H_2O}$) of water molecules surrounding ATI($H^+$) (Fig. 9A). One can see the following trends: a positive charge $Q_{N2}$ on the nitrogen atom N2 gradually increases with the rise of $n_{H_2O}$, whereas a negative charge $Q_{O1}$ on the oxygen atom O1 becomes smaller. The nitrogen atom N4 of the imidazoline ring acquires additional negative charge. Computations with the use of two different basis sets (B1 or B2) gave

---

molecules in ($H_2O$)$_{27}$ cluster is characterised by the energy decrease $\Delta E_{H_2O} \cong -13.5$ kcal/mol per one water molecule, which is the same by the order of magnitude as the water evaporation energy (will be published elsewhere).

similar results (Fig. 9A, Table 4). Note that the water-induced redistribution of charges is determined mainly by polarisation effects rather than due changes in ATI geometry. Actually, when we considered the "naked" ATI($H^+$) molecules (without water), which conformations were fixed so to imitate the water-induced structural changes in ATI($H^+$) geometry, computed partial charges $Q_{O1}$, $Q_{N2}$ and $Q_{N4}$ remained practically unchanged (Fig. 9B).

### 3.4.2. Electron spin density on atoms.

Figure 8 presents the distribution of spin densities over the atoms of protonated and deprotonated ATI molecules in the gas phase or surrounded by 41 water molecules (here we show spin densities on atoms greater than 0.01 a.u.). Calculated electron spin density is localised mainly in the nitroxide fragment N-O, whereas spin densities on C3 and C6 atoms of the imidazoline ring and methyl groups are insignificant. One can also see from Fig. 8 that the redistribution of electron spin density induced by protonation and/or hydration of a spin label molecule occurs mainly in the N-O bond. Therefore, we consider below in more details only the nitroxide fragment of a spin label.

*Effects of protonation*. Fig. 8 shows that ATI protonation at the N4 position leads to a certain decrease in the spin atom density on N2 atom (parameter $\sigma_{N2}$) at expense of the spin density on the oxygen atom O1 (parameter $\sigma_{O2}$). For ATI in the gas phase, the total spin density in the nitroxide fragment remains practically unchanged after ATI protonation ($\sigma_{N-O} = \sigma_{N2} + \sigma_{O2} = 0.86$). In hydrated ATI, parameter $\sigma_{N-O}$ slightly decreases (from 0.83 to 0.81) with ATI protonation.

Distribution of spin density in nitroxide radicals is usually interpreted qualitatively in terms of the two main resonance structures **I** and **II** (Scheme 1) which differ with respect to localisation of the unpaired electron on O1 and N2 atoms. A decrease in the $\sigma_{N2}$ value caused by the protonation of the nitrogen atom of the imidazoline ring could be explained by stabilisation of the structure **I** due to the rise of so called "electron withdrawing power" of the N4 atom[40]. This explanation is consistent with the fact that protonation of the N4 atom reduces its negative charge (Fig. 7). This, in turn, reduces the contribution of the dipolar resonance structure **II**, thus resulting in the decrease in a spin density on the N2 atom.

*Effects of hydration*. The hydration of both deprotonated and protonated forms of ATI molecule causes a rise in a spin density on the N2 atom and concomitant decrease in $\sigma_{O2}$ value. A partial displacement of spin density from the O1 atom towards the N2 atom can be explained by water-induced stabilisation of the dipolar structure **II** at expense of the non-polar structure **I**. Below we consider the hydration effects in more details, focusing our attention, for certainty, on the protonated form of a spin label, ATI($H^+$).

Fig. 10 shows the plots of parameters $\sigma_{N2}$ and $\sigma_{O2}$ versus a number of water molecules surrounding ATI($H^+$), demonstrating the water-induced shift of spin density towards the N2 atom with a gradual hydration of a spin label. Calculations performed at two levels of theory (PBE/B1 and PBE/B2//B1) led us to similar results. Certain irregularities in the plots of $\sigma_{N2}$ and $\sigma_{O1}$ observed in the interval $10 \leq n_{H_2O} \leq 35$ could arise from the system discreteness, when a water shell around a spin label molecule is not yet high enough to be considered as a complete hydration sphere. The total spin density in the N-O fragment ($\sigma_{N-O} = \sigma_{N2} + \sigma_{O2}$) slightly decreases ($\leq 0.03$-0.05 a.u.) with a spin label hydration. According to our calculations, a minor spin density (ac. 0.02 a.u.) appears on two water molecules situated in the nearest vicinity of the N-O fragment. Note that a water-induced increase in spin density $\sigma_{N2}$ on the nitrogen atom strongly correlates with a rise of its charge $Q_{N2}$ (Fig. 11A).

Similarly to charge plot (Fig. 9B), spin density $\sigma_{N2}$ does not change with those variations of ATI(H$^+$) geometry that imitated the water-induced structural changes (Fig. 11A). This suggests that a partial shift of spin density from the oxygen atom O1 towards the nitrogen atom N2 is determined mainly by water-induced polarisation effect rather than by structural distortions of ATI(H$^+$) molecule. We would also like to stress that calculations performed at different levels theory (Table 4) lead to similar conclusions about the protonation and hydration effects on spin density distribution in the N-O bond.

### 3.4.3. Isotropic splitting constant $a_{iso}$(N2).

For nitroxide radicals, the isotropic hyperfine splitting constant $a_{iso}$(N2) is determined by spin density $\rho(\vec{r}_N)$ at the nitrogen nucleus[41]. Similarly to $\sigma_{N2}$ (Fig. 11A), the $a_{iso}$(N2) value, which is proportional to calculated parameter $\rho(\vec{r}_N)$, increases with the ATI(H$^+$) hydration (Table 5, Fig. 11B). However, contrary to spin density on atoms, an accurate theoretical determination of spin density at nuclei is very difficult task. This is because calculated spin densities at nuclei are extremely sensitive to a choice of the computation method (exchange-correlation functional and basis set). Therefore, we have to compare experimental data with theoretical values of $a_{iso}$(N2) computed at different levels of theory.

According to EPR data[9], deprotonated ATI molecule in water is characterized by the splitting constant $a_{iso}$ = 15.8 G, while protonated form of ATI has $a_{iso}$ = 15.0 G. Experimental hyperfine splitting constants for gas phase ATI are not available. As one of approximations to evaluation of gas phase constants, one could take the $a_{iso}$ values measured for spin labels dissolved in organic solvents, i.e., in CCl$_4$. The comparison of EPR spectra of a variety of different spin labels dissolved in water and CCl$_4$ (see for references[1,7]) shows that the difference $\Delta a_{iso} = a_{iso}$(H$_2$O) - $a_{iso}$(CCl$_4$) falls in the interval from 1 G to 2 G. Bearing in mind these data, it is reasonable to propose that in the gas phase the hyperfine splitting constant should be about of $a_{iso}$(gas) $\approx$ 13-14 G.

Table 5 lists the $a_{iso}(N2)$ constants calculated with the use of different functionals and basis sets. One can see that $a_{iso}(N2)$ determined at the levels PBE/B2//B1 or BLYP/B2//B1, which have been used in most of our calculations, are closer to experimental constants than that computed with the use of B1 basis. For ATI and ATI(H$^+$) in the gas phase, calculations at the level BLYP/B2//B1 give $a_{iso}(N2)$ = 11.2 G and $a_{iso}(N2)$ = 11.5 G, respectively. Smaller values, 10.6 and 10.5, were obtained at PBE/B2//B1 level of computations. Calculations with the use of basis B1 lead to substantially underestimated values of $a_{iso}(N2)$.

Note that our calculations of $a_{iso}(N2)$ for gas phase ATI and ATI(H$^+$) with GAUSSIAN98 package and the B3LYP functional, which is one of the most popular approximations in DFT computations, gave scattered results depending on the basis set[42]. Thus, the use of conventional approaches to calculations of $a_{iso}(N2)$, e.g., B3LYP/6-31G or B3LYP/EPR-II, does not give practical advantages over the calculation methods used in this work. In the meantime, the program PRIRODA[35] allows the low cost computations to be performed without the loss of computation accuracy for such complex systems as that used in our work.

Fig. 11B shows the plot of $a_{iso}(N2)$ versus a number of surrounding water molecules (calculations at PBE/B2//B1 level). Parameter $a_{iso}(N2)$ non-monotonously change with gradual expansion of a water shell around ATI(H$^+$). When an aqueous sphere around ATI(H$^+$) becomes complete ( $n_{H_2O}$ = 41), the water-induced rise of $a_{iso}(N2)$ comprises $\Delta a_{iso}(N2) \approx 2.7$ G, which is typical of experimental difference between the hyperfine splitting constants measured for various nitroxide radicals dissolved in water and nonpolar solvents[1,7]. Comparing the results obtained with different functionals for hydrated ATI molecules (Table 5), one can see that functionals BLYP and PBE provide underestimated by 1.5-1.8 G values of $a_{iso}(N2)$. Concerning certain underestimations in computations of $a_{iso}(N2)$ with PBE and BLYP functionals, we would like to note that there is an obvious trend of increasing the $a_{iso}(N2)$ with the rise of $n_{H_2O}$. Therefore, one might expect to obtain realistic values of $a_{iso}(N2)$ if more water molecules were involved into an aqueous sphere around ATI. Further improvements of calculated $a_{iso}(N2)$ might occur due to proper corrections for the vibrational modulations of $a_{iso}(N2)$. For instance, according to[40], in 2,2,4,4-tetramethyl substituted nitroxides $\Delta a_{iso}(N2)^{vibr} \approx 1$ G.

It was also temping to simulate the effect of an aqueous environment by using the IPCM approach. However, results of our calculations (Table 6) clearly demonstrate that the IPCM approach does not help when one wants to evaluate the influence of a solvent polarity on electronic characteristics of ATI. In both cases, deprotonated and protonated ATI, the IPCM approach gives practically the same values of $a_{iso}(N2)$ as in the gas phase. Thus, the IPCM method cannot help in simulation of a water-induced rise of $a_{iso}(N2)$ that was, however, successfully computed by means of the obvious consideration of a water environment.

## CONCLUDING REMARKS

The main question of this work has been formulated as follows: how many water molecules should be considered for an adequate simulation of the first hydration sphere around a spin label ATI? Investigating how structural and energy parameters of a cluster ATI(H$^+$)–(H$_2$O)$_n$ depend on the number of water molecules, we came to the conclusion that a water cluster of $n_{H_2O} \geq 40$ water molecules can be considered as an appropriate model for a complete aqueous shell around ATI. In a reasonable agreement with experimental data, calculated electron spin density on the nitrogen atom of the nitroxide fragment increased with a stepwise surrounding of a spin label molecule with water. Hydration of a spin label caused certain changes in its geometric parameters - bond lengths and angles. However, it was demonstrated, it is the water-induced polarisation effect rather than changes in a spin label geometry that caused the increase in the isotropic splitting constant $a_{iso}$(N2).

Calculating the energy characteristics of ATI molecule surrounded by water, we came to the conclusion that ATI protonation occurs at the nitrogen atom of the imidazoline ring rather than at the amine group. Having the energy diagram for hydrated ATI molecule (Fig. 3B, compare energy levels for states ATI(H$^+$)–H$_3$O$^+$(H$_2$O)$_{41}$ and ATI–H$_3$O$^+$(H$_2$O)$_{40}$), it might be interesing to evaluate the pK value of ATI. However, a correct solution of this task is obscured by significant scattering of calculated energy levels, which depend, in particular, on the H$_3$O$^+$ ion position in the hydrogen-bonded network of water molecules around ATI. According to our calculations, the energy fluctuations could reach ac. 10 kcal/mol (detailed analysis of energy fluctuations in this system will be published elsewhere).

Computations of spin density partitioning on the atoms of deprotonated and protonated ATI molecules in the gas phase or surrounded by 41 water molecules (Fig. 8) are in a fair agreement with experimental data concerning the changes in the hyperfine splitting constant with ATI protonation and hydration. Our calculations of the isotropic hyperfine constant predicted a water-induced increase in $a_{iso}$(N2) as $\Delta a_{iso}$(N2) = 2-2.5 G, which is typical of nitroxide radicals. However, the absolute $a_{iso}$(N2) values calculated for hydrated ATI molecule is lower (ac. 1.8-2.6 G, depending on the computation level) than experimentally measured constant $a_{iso}$(N2) = 15.0 G (Table 5). Certain shortcomings in computations of $a_{iso}$(N2) arise from a high sensitivity of calculated spin densities at nuclei to a choice of the basis set. One might expect that proper corrections for the vibrational modulations of $a_{iso}$(N2), as well as a consideration of a greater number of water molecules surrounding ATI, should lead to much more accurate values of calculated $a_{iso}$(N2). We hope that further progress in computational chemistry will help to overcome the problem of accurate calculations of EPR parameters of spin labels.

**Acknowledgements**

The authors acknowledge Dr. D.N. Laikov for his program PRIRODA and useful discussions. We also thank D.V. Besedin for valuable technical assistance. The authors acknowledge funding from INTAS (grants 99-1086 and 01-483) and the Russian Foundation for Basic Researches (grant 03-04-48981).

**Table 1**. The examples of basis sets B1 and B2 (orbital and auxiliary) used in this work.

| Basis set | Atoms | Orbital basis set | Auxiliary basis set |
|---|---|---|---|
| B1 | C,O,N | (11s6p2d)/[6s3p2d] | (10s3p3d1f) |
| | H | (5s1p)/[3s1p] | (5s2p) |
| B2 | C,O,N | (17s4p1d)/[3s2p1d] | (25s11p6d1f)/[4s3p2d1f] |
| | H | (8s1p)/[2s1p] | (8s1p)/[2s1p] |

**Table 2**. Geometry parameters of 4-hydroxy-2,2,6,6-tetramethylpiperidine-1-oxyl (TEMPOL) in gas phase computed at the PBE/B1 and BLYP/B1 levels and their comparison with the literature data. Calculations were performed for two possible conformations of TEMPOL, conformer 1 and conformer 2, which differ with respect to orientation of the OH group relatively to the ring plane.

| | PBE/B1 conformer 1 | BLYP/B1 conformer 1 | PBE/B1 conformer 2 | BLYP/B1 conformer 2 | Experim.[39] | B3LYP/6-31G* [39] |
|---|---|---|---|---|---|---|
| Bond length, Å | | | | | | |
| O1-N2 | 1.287 | 1.288 | 1.286 | 1.300 | 1.291 | 1.285 |
| N2-C3 | 1.509 | 1.508 | 1.510 | 1.527 | 1.498 | 1.503 |
| C3-C4 | 1.545 | 1.542 | 1.543 | 1.551 | 1.526 | 1.542 |
| C4-C5 | 1.527 | 1.527 | 1.526 | 1.532 | 1.517 | 1.528 |
| C5-O6 | 1.441 | 1.444 | 1.438 | 1.456 | 1.422 | 1.426 |
| O6-H | 0.971 | 0.973 | 0.971 | 0.974 | 0.895 | 0.971 |
| Bond angle, grad | | | | | | |
| O1-N2-C3 | 116.0 | 115.9 | 115.9 | 115.8 | 116.2 | 115.6 |
| C3-N2-C3 | 123.8 | 123.8 | 123.9 | 124.3 | 125.4 | 124.7 |
| N2-C3-C4 | 110.3 | 110.2 | 110.3 | 110.0 | 110.1 | 110.1 |
| C3-C4-C5 | 116.3 | 116.3 | 113.8 | 114.1 | 113.1 | 114.1 |
| C4-C5-C4 | 109.7 | 109.8 | 108.6 | 109.2 | 108.3 | 109.1 |
| C4-C5-O6 | 111.4 | 111.4 | 111.7 | 111.5 | 112.3 | 111.7 |
| C5-O6-H | 107.5 | 107.2 | 107.7 | 107.8 | 104.3 | 107.4 |
| Torsion angle, grad | | | | | | |
| O1-N2-C3-C4 | 168.3 | 168.3 | -170.3 | -170.8 | -167.5 | -169.7 |
| N2-C3-C4-C5 | 43.5 | 43.5 | -45.7 | -44.8 | 44.6 | 44.9 |
| C3-C4-C5-O6 | 72.4 | 73.0 | 177.3 | 177.1 | 173.3 | 177.3 |
| C4-C5-O6-H | 179.9 | 178.4 | 59.4 | 61.4 | 61.1 | 61.2 |

**Table 3.** Calculated geometry parameters of 4-amino-2,2,5,5-tetramethyl-3-imidazoline-N-oxyl molecule in deprotonated (ATI) and protonated (ATI(H$^+$)) forms in a gas phase or surrounded by 41 water molecules. Computations were performed at the PBE/B1 level.

| | ATI | ATI(H$^+$) | ATI-(H$_2$O)$_{41}$ | ATI(H$^+$)-(H$_2$O)$_{41}$ |
|---|---|---|---|---|
| Distance, Å | | | | |
| O1-N2 | 1.275 | 1.267 | 1.285 | 1.281 |
| N2-C3 | 1.507 | 1.493 | 1.495 | 1.488 |
| N2-C6 | 1.487 | 1.498 | 1.493 | 1.485 |
| C3-N4 | 1.457 | 1.488 | 1.452 | 1.472 |
| N4-C5 | 1.290 | 1.321 | 1.313 | 1.326 |
| C5-C6 | 1.524 | 1.513 | 1.534 | 1.518 |
| C5-N7 | 1.376 | 1.327 | 1.341 | 1.314 |
| N4-H | - | 1.267 | - | 1.039 |
| Angle, grad | | | | |
| C6-N2-C3 | 111.7 | 114.7 | 111.3 | 113.9 |
| N2-C3-N4 | 103.5 | 98.5 | 103.0 | 98.7 |
| C3-N4-C5 | 110.0 | 115.3 | 109.7 | 114.3 |
| N4-C5-C6 | 116.9 | 111.8 | 115.5 | 111.7 |
| N2-C6-C5 | 97.6 | 99.1 | 97.1 | 98.3 |
| Torsion angle, grad | | | | |
| O1-N2-C3-N4 | 176.9 | -174.1 | 164.0 | -174.8 |
| O1-N2-C6-C5 | -176.4 | 174.4 | -168.3 | 175.5 |
| O1-N2-C6-C3 | -172.3 | 166.7 | 176.6 | -167.2 |
| N2-C3-N4-C5 | -3.2 | 3.9 | 15.0 | -10.8 |
| N2-C6-C5-N4 | 2.2 | -4.8 | -5.7 | 9.8 |
| C3-N4-C5-C6 | 0.7 | 0.6 | -6.1 | 0.7 |

**Table 4.** Spin densities ($\sigma_{O1}$ and $\sigma_{N2}$) and partial charges ($Q_{O1}$, $Q_{N2}$, $Q_{N3}$) on atoms O1, N2 and N4 calculated for deprotonated (ATI) and protonated (ATI(H+)) forms of 4-amino-2,2,5,5-tetramethyl-3-imidazoline-N-oxyl in gas phase or surrounded by 41 water molecules using the different methods.

| System | Functional/Basis | $\sigma_{O1}$ | $\sigma_{N2}$ | $Q_{O1}$ | $Q_{N2}$ | $Q_{N4}$ |
|---|---|---|---|---|---|---|
| ATI | PBE/B1 | 0.459 | 0.397 | -0.216 | 0.062 | -0.198 |
| | PBE/B2//B1 | 0.476 | 0.388 | -0.206 | 0.076 | -0.196 |
| | BLYP/B2//B1 | 0.479 | 0.395 | -0.209 | 0.073 | -0.199 |
| ATI(H$^+$) | PBE/B1 | 0.487 | 0.370 | -0.153 | 0.066 | -0.035 |
| | PBE/B2//B1 | 0.503 | 0.363 | -0.145 | 0.081 | -0.020 |
| | BLYP/B2//B1 | 0.508 | 0.368 | -0.149 | 0.077 | -0.023 |
| ATI-(H$_2$O)$_{41}$ | PBE/B1 | 0.379 | 0.447 | -0.142 | 0.088 | -0.165 |
| | PBE/B2//B1 | 0.392 | 0.443 | -0.138 | 0.103 | -0.161 |
| | BLYP/B2//B1 | - | - | - | - | - |
| ATI(H$^+$)-(H$_2$O)$_{41}$ | PBE/B1 | 0.373 | 0.438 | -0.116 | 0.096 | -0.054 |
| | PBE/B2//B1 | 0.383 | 0.435 | -0.110 | 0.110 | -0.039 |
| | BLYP/B2//B1 | 0.380 | 0.446 | -0.120 | 0.106 | -0.046 |

**Table 5.** Computed values of the isotropic hyperfine splitting constant $a_{iso}(N2)$ calculated for deprotonated (ATI) and protonated (ATI(H+)) forms of 4-amino-2,2,5,5-tetramethyl-3-imidazoline-N-oxyl in gas phase or surrounded by 41 water molecules.

| System | $a_{iso}(N2)$ (G) | | |
|---|---|---|---|
| | PBE/B1 | PBE/B2//B1 | BLYP/B2//B1 |
| ATI | 7.0 | 10.6 | 11.2 |
| ATI(H$^+$) | 7.2 | 10.5 | 11.5 |
| ATI$-$(H$_2$O)$_{41}$ | 8.5 | 12.8 | 13.5 |
| ATI$-$H$_3$O$^+$(H$_2$O)$_{40}$ | 8.8 | 13.2 | - |
| ATI(H$^+$)$-$(H$_2$O)$_{41}$ | 9.0 | 13.2 | 13.5 |

**Table 6.** Spin densities on the N2 nucleus calculated for deprotonated and protonated forms of ATI in gas phase and aqueous shell and their comparison with that values computed within the IPCM approach (dielectric constant of surrounding medium $\varepsilon = 78.39$). Computation levels PBE/B1 and PBE/B2//B1.

| System | $a_{iso}(N2)$ (G) | | | |
|---|---|---|---|---|
| | PBE/B2//B1 | PBE/B2//B1 (IPCM) | PBE/B1 | PBE/B1 (IPCM) |
| ATI | 10.6 | 10.6 | 7.0 | 7.0 |
| ATI·H$^+$ | 10.5 | 10.5 | 7.2 | 7.2 |
| ATI·H$^+$$-$(H$_2$O)$_5$* | 11.1 | 11.1 | 7.1 | 7.1 |

\* The data presented correspond to ATI$-$(H$_2$O)$_5$ cluster shown on Fig.2.

Scheme 1.

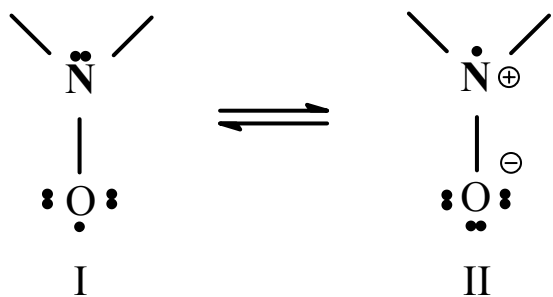

I             II

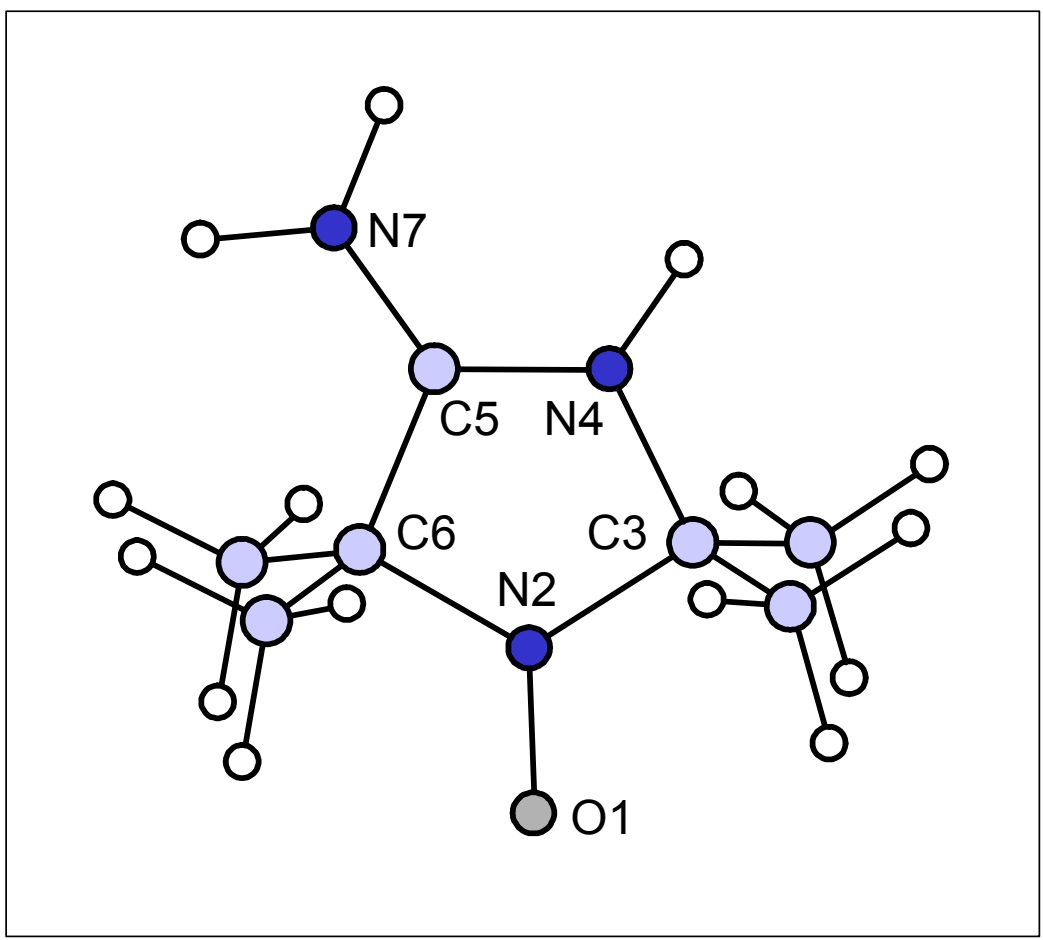

**Fig. 1.** Chemical structure and numbering of 4-amino-2,2,5,5-tetramethyl-3-imidazoline-N-oxyl (ATI).

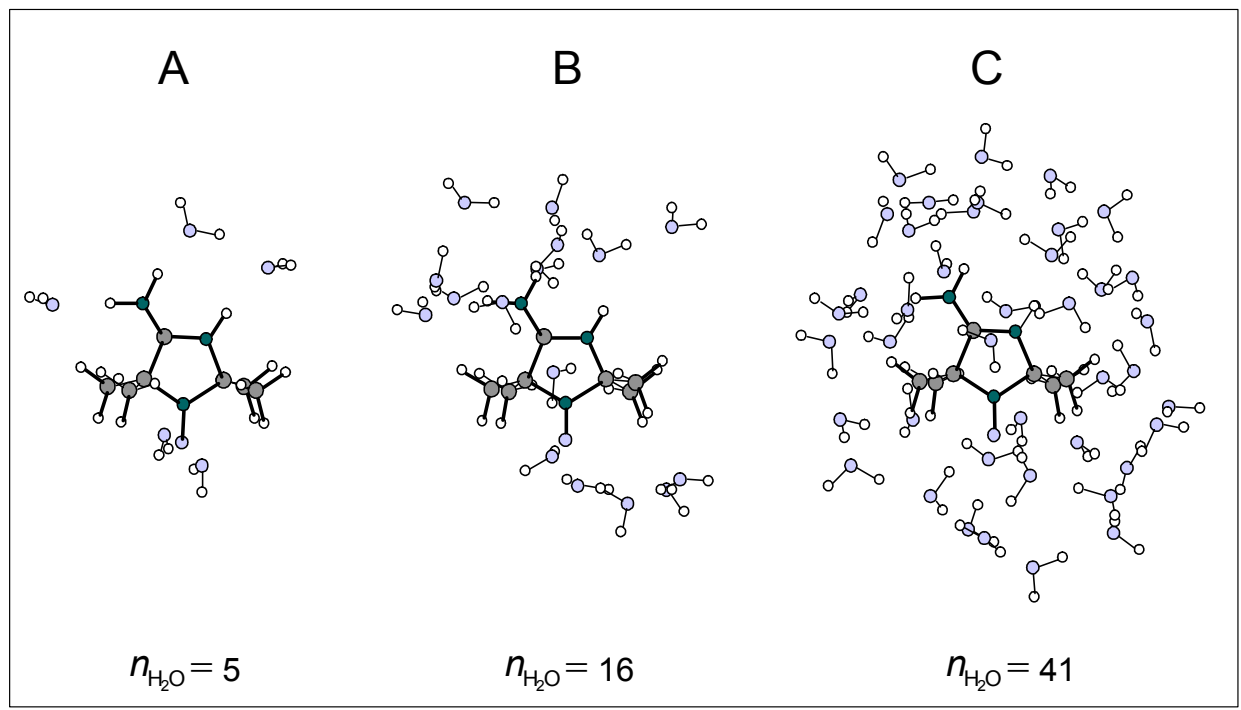

**Fig. 2.** Chemical structures of protonated 4-amino-2,2,5,5-tetramethyl-3-imidazoline-N-oxyl, ATI(H$^+$), surrounded by water molecules. **A** - ATI(H$^+$) surrounded by five H$_2$O molecules; **B** - ATI(H$^+$) surrounded by 17 H$_2$O molecules; **C** - ATI(H$^+$) surrounded by 41 H$_2$O molecules.

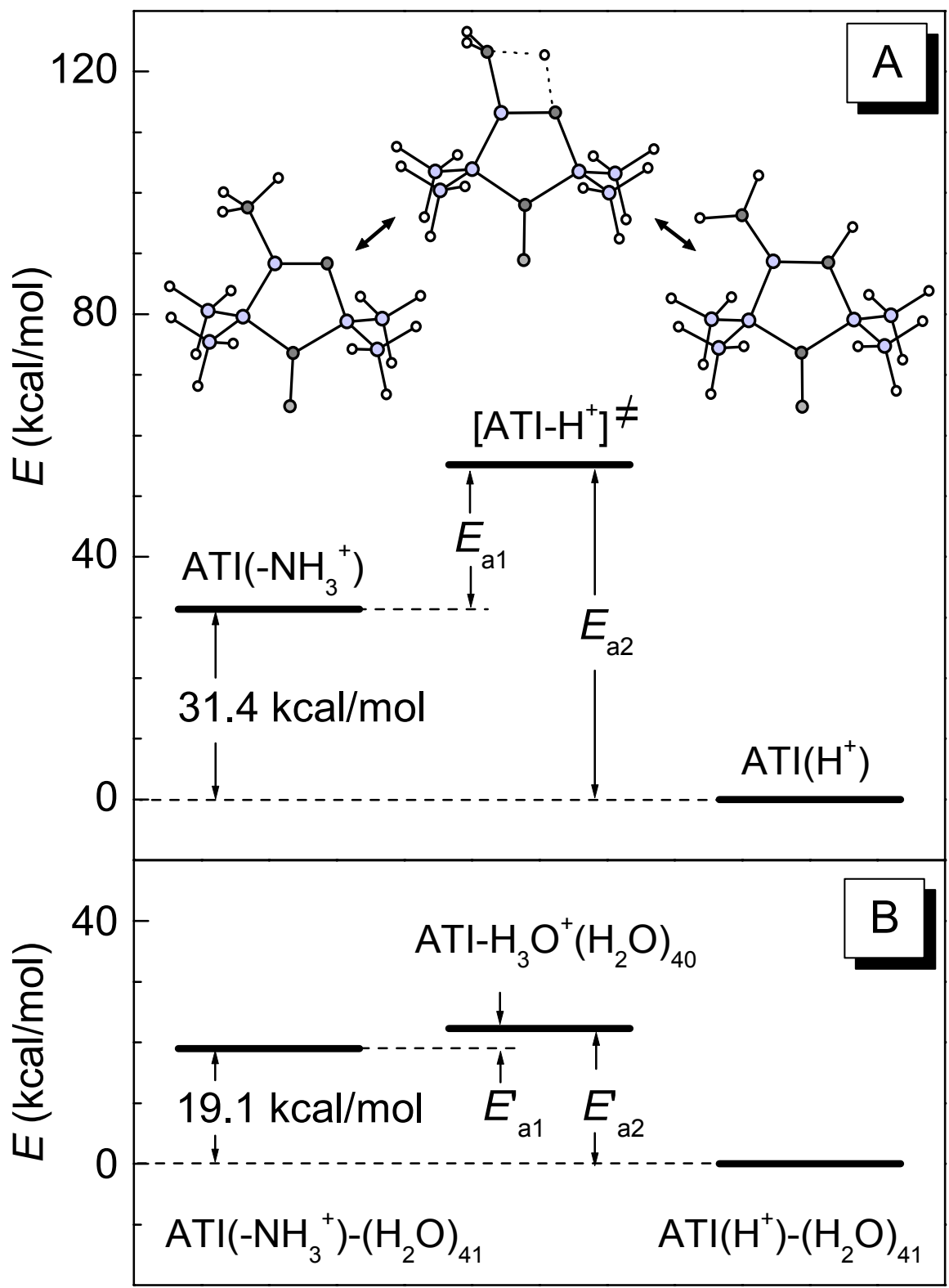

**Fig. 3.** Energy diagrams characterising proton transfer between the amino group and the ring nitrogen N4 of ATI moleclule. **A** – ATI in gas phase; top figures show calculated structures of ATI molecule protonated either at the amino group (left) or at the ring nitrogen atom N4 (right), central figure depicts the transient state of protonated ATI. **B** – ATI molecule surrounded by 41 $H_2O$ molecules. ATI-$H_3O^+$($H_2O$)$_{40}$ depicts the model, in which the proton is replaced from the amino group to the nearest water molecule of the hydration sphere.

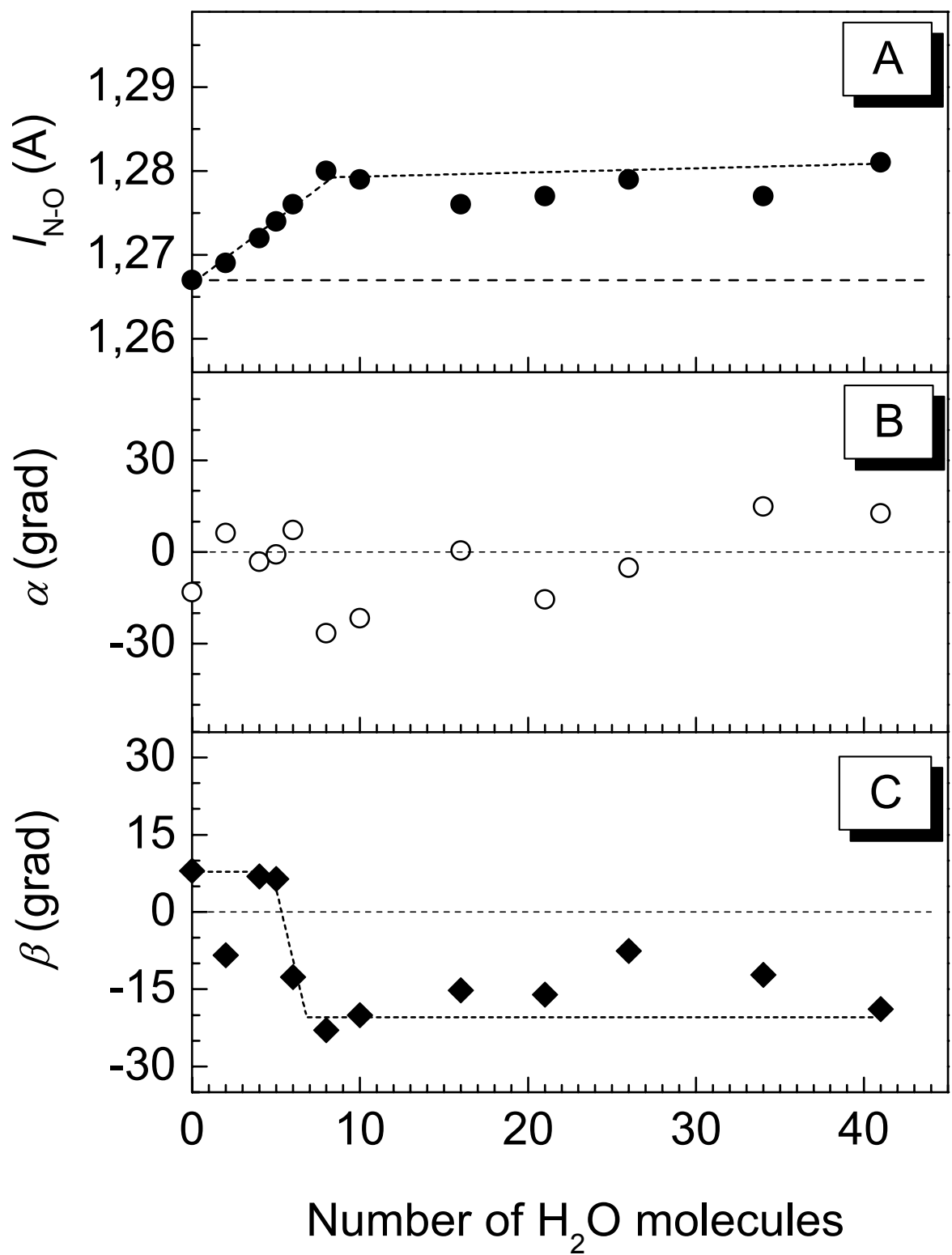

**Fig. 4.** Plots of structural parameters $l_{N-O}$ (A), $\alpha$ (B) and $\beta$ (C), characterising geometry of protonated molecule $ATI(H^+)$, versus a number of water molecules surrounding $ATI(H^+)$.

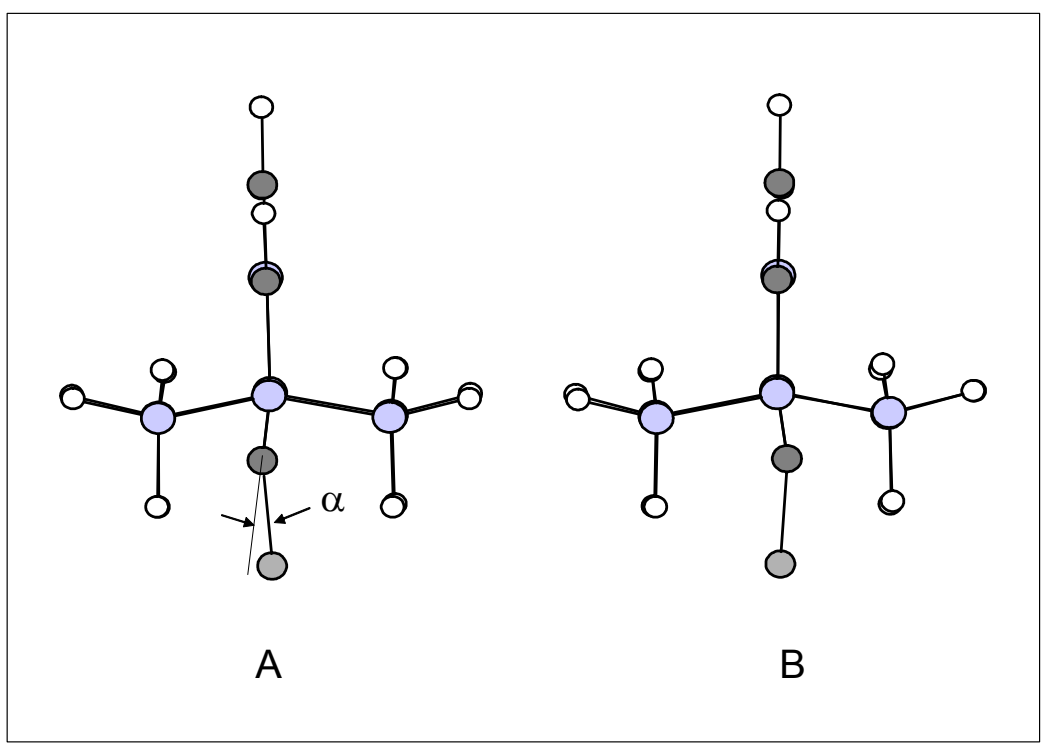

**Fig. 5.** Calculated structures of two conformers of protonated ATI molecule in the gas phase.

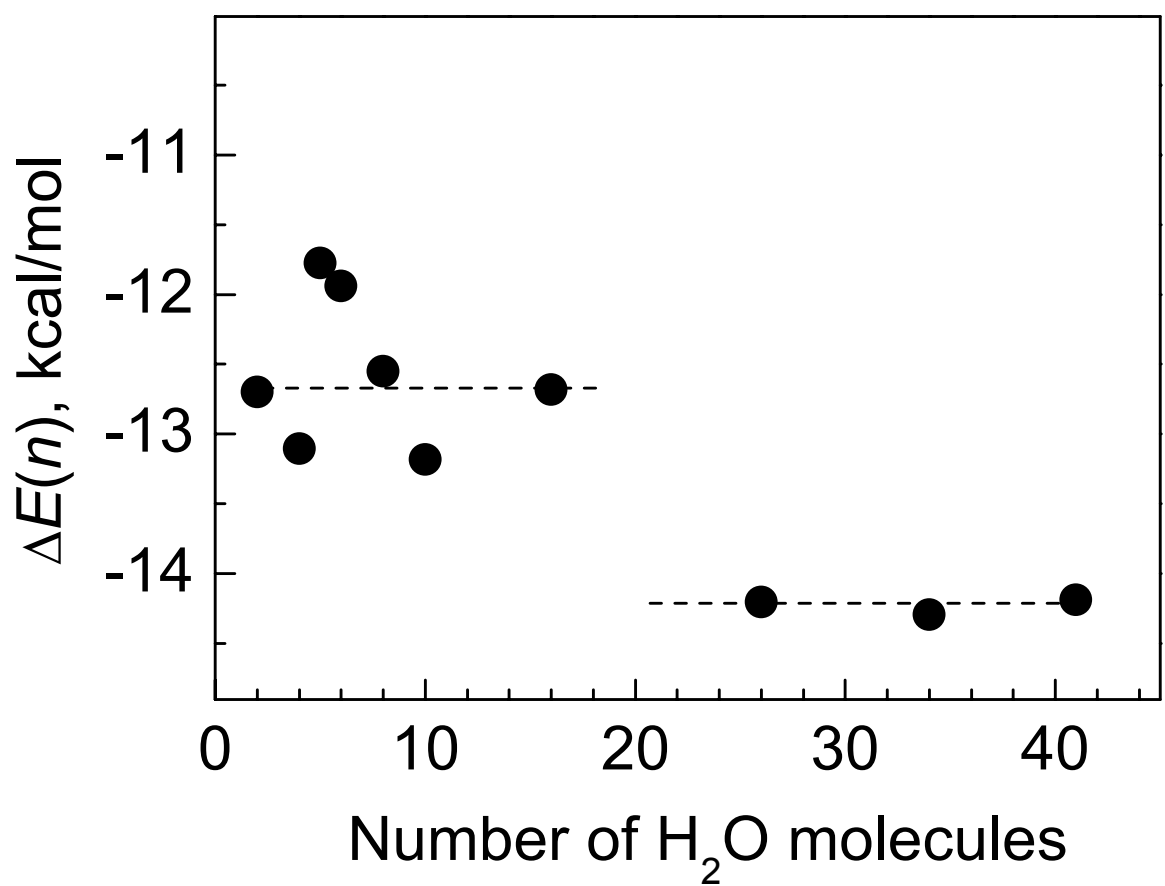

**Fig. 6.** Energy parameter $\Delta E(n)$ versus a number of $H_2O$ molecules surrounding protonated ATI($H^+$) molecule (see text for details).

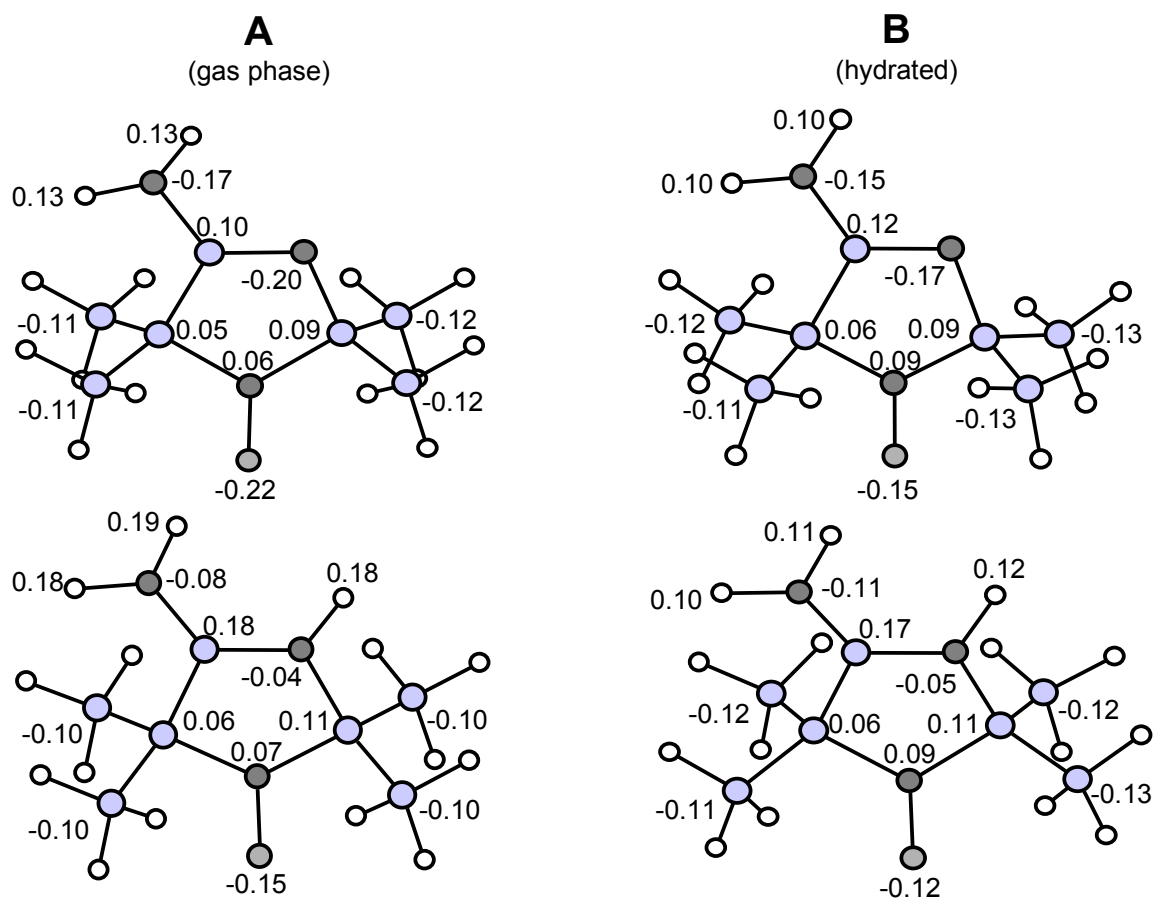

**Fig. 7.** Electric charges on atoms of deprotonated (top pictures) and protonated (bottom pictures) ATI molecules. **A** – spin labels in the gas phase. **B** – spin labels surrounded by 41 water molecules.

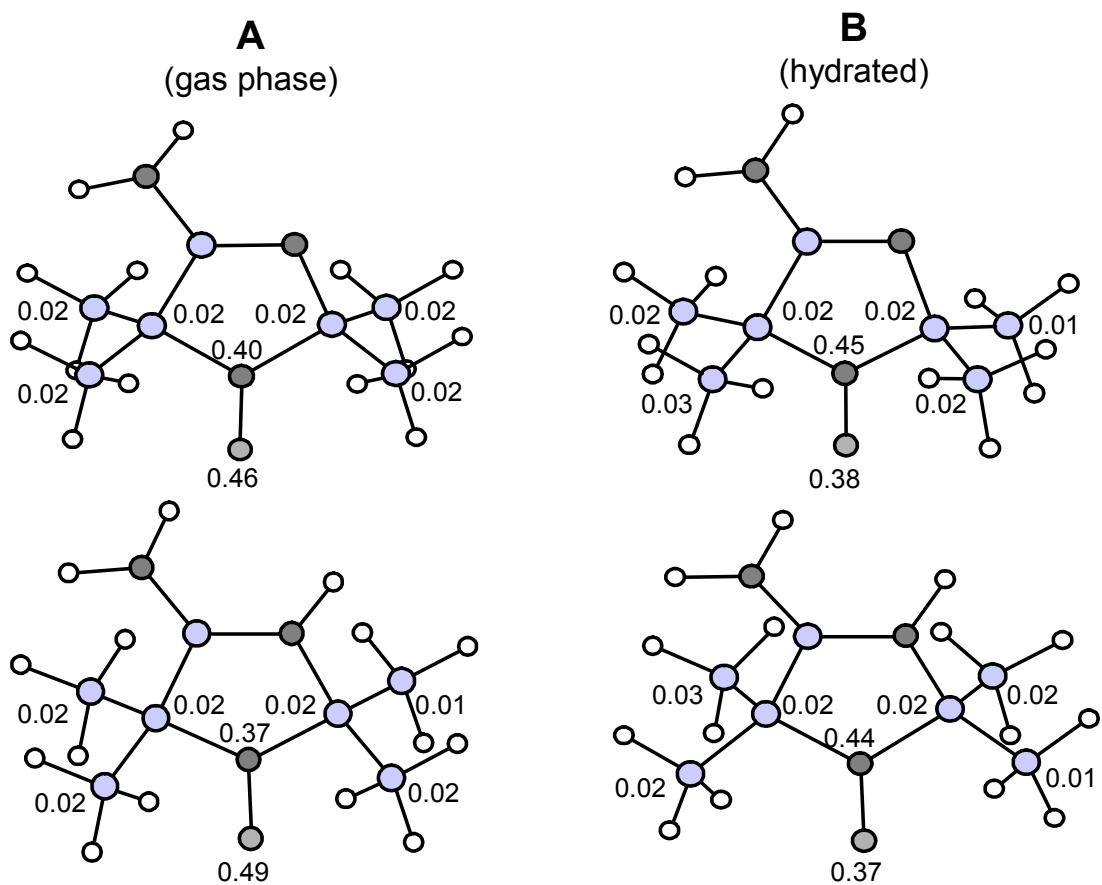

**Fig. 8.** Spin densities on atoms of ATI (top pictures) and ATI(H$^+$) (bottom pictures) molecules. **A** - spin labels in the gas phase. **B** - spin labels surrounded by 41 water molecules.

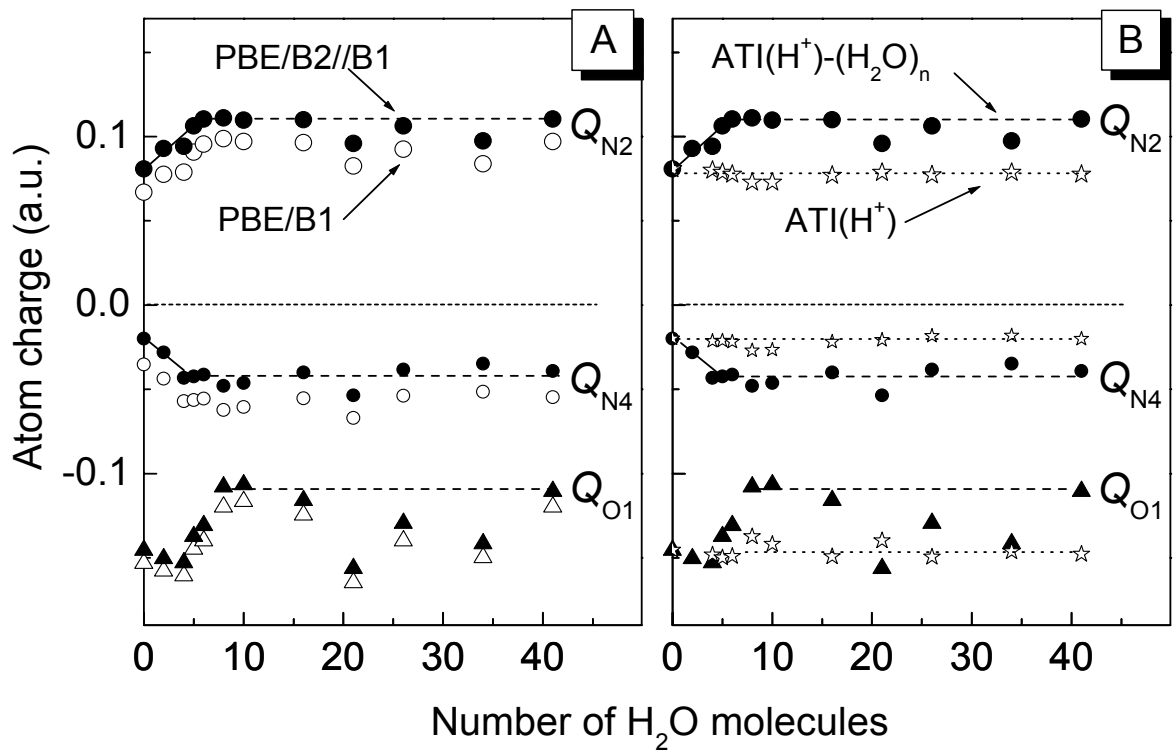

**Fig. 9.** Plots of partial charges on atoms O1 ($Q_{O1}$), N2 ($Q_{N2}$) and N4 ($Q_{N4}$) in protonated ATI(H$^+$) molecule versus a number of water molecules surrounding ATI(H$^+$). **A** – comparison of results calculated at the levels PBE/B2//B1 (solid symbols) or PBE/B1 (open symbols). **B** – partial charges calculated (PBE/B2//B1 level of theory) for ATI(H$^+$) surrounded by water (solid symbols) or "naked" ATI(H$^+$) in the gas phase (open stars). In the latter case, calculations were performed for ATI(H$^+$) with fixed conformations that corresponded to ATI(H$^+$) molecules surrounded by various numbers of water molecules (see text for details).

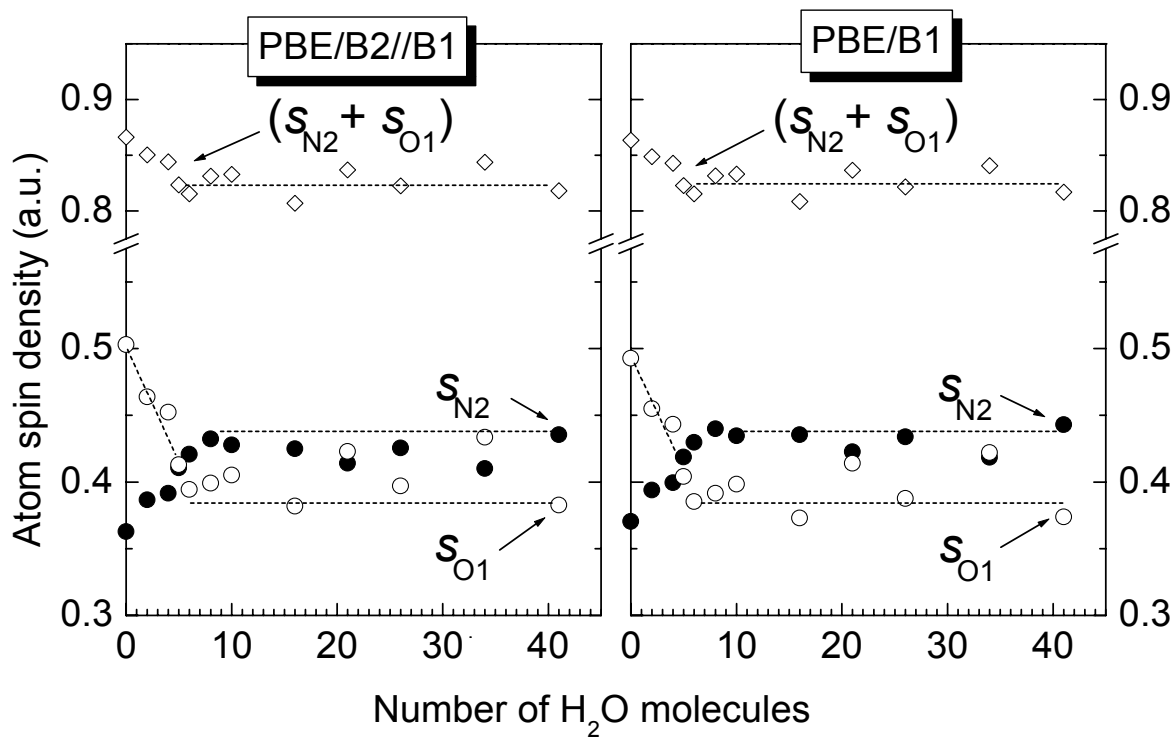

**Fig. 10.** Plots of spin densities on atoms O1 ($\sigma_{O1}$), and N2 ($\sigma_{N2}$) in ATI(H$^+$) versus a number of H$_2$O molecules in a cluster ATI(H$^+$)–(H$_2$O)$_n$. **A** – computations at PBE/B2//B1 level. **B** - computations at PBE/B1 level. In the top of the picture the sum ($\sigma_{O1} + \sigma_{N2}$) is presented.

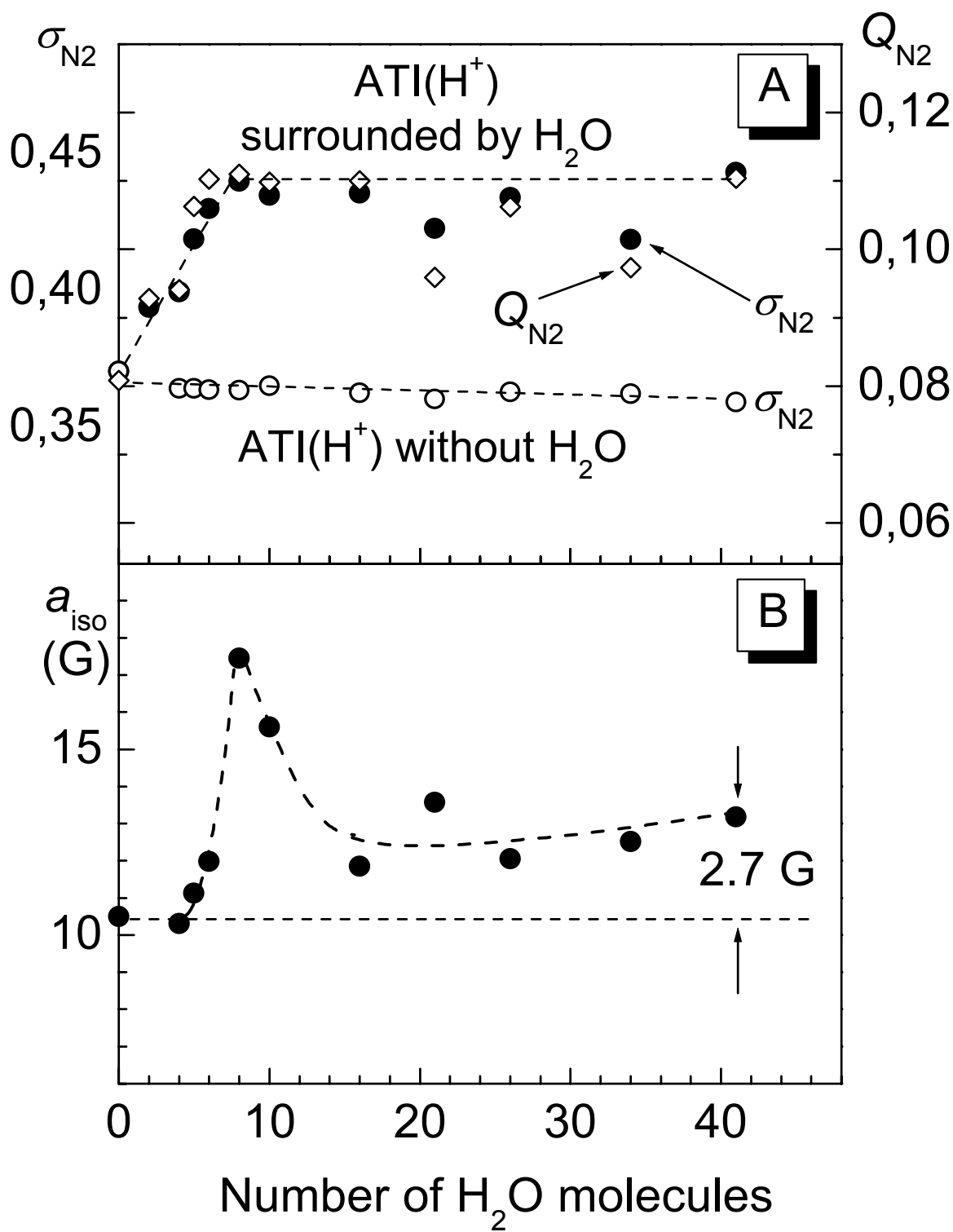

**Fig. 11. A** - Plots of spin density $\sigma_{N2}$ (solid circles) and electric charge $Q_{N2}$ (open diamonds) on N2 atom versus a number of $H_2O$ molecules surrounding ATI($H^+$). Open circles correspond to $\sigma_{N2}$ calculated for ATI($H^+$) molecules ("naked" ATI($H^+$)) in gas phase with fixed conformations corresponding to that of ATI($H^+$) molecules surrounded by different numbers of water molecules.

**B** - hyperfine splitting constant $a_{iso}$ calculated versus a number of $H_2O$ molecules surrounding ATI($H^+$). Calculations at PBE/B2//B1 level.